\documentstyle[prl,twocolumn,aps]{revtex}
\def\tri{{{}^3{\rm H}}}
\def\hel{{{}^3{\rm He}}}
\def\het{{{}^3{\rm He}}}
\def\heq{{{}^4{\rm He}}}

\def\be{\begin{equation}}
\def\ee{\end{equation}}
\def\bea{\begin{eqnarray*}}
\def\eea{\end{eqnarray*}}
\def\ra{\rightarrow}

\def\bi{\begin{itemize}}
\def\ei{\end{itemize}}
\def\qn{{J}}
\def\qnz{{0}}

\begin{document}
\draft

\wideabs{
\title{The $A_y$ Problem for {\boldmath $p-\het$} Elastic Scattering} 
\author{
M.\ Viviani$^{({\rm a})}$,
A.\ Kievsky$^{({\rm a})}$,
S.\ Rosati$^{({\rm a,b})}$,
E.A.\ George$^{({\rm c,d})}$, and
L.D.\ Knutson$^{({\rm d})}$}
\address{$^{({\rm a})}$INFN, Sezione di Pisa, I-56100 Pisa, Italy}
\address{$^{({\rm b})}$Department of Physics, University of Pisa, I-56100 Pisa, Italy}
\address{$^{({\rm c})}$Physics Department, Wittenberg University,
                       Springfield, Ohio 45501}
\address{$^{({\rm d})}$Physics Department, University of Wisconsin,
                       Madison, Wisconsin 53706}
\date{\today}
\maketitle
\begin{abstract}
We present evidence that numerically accurate quantum calculations
employing modern internucleon forces do not reproduce the proton
analyzing power, $A_y$, for $p-\het$ elastic scattering at low
energies.  These calculations underpredict new measured
analyzing powers by approximately 30\% at $E_{c.m.} = 1.20$ MeV and by
40\% at $E_{c.m.} = 1.69$ MeV, an effect analogous to a well-known
problem in $p$-$d$ and $n$-$d$ scattering.
The calculations are performed using the complex Kohn variational
principle and the (correlated) Hyperspherical Harmonics technique
with full treatment of the Coulomb force.
The inclusion of the three-nucleon interaction does not improve the
agreement with the experimental data.
\end{abstract}


}

Over the past decade, numerical calculations for three-nucleon
systems have reached a high degree of precision. This has made it
possible to carry out accurate quantum mechanical computations for
a variety of processes including 1) $N$-$d$ elastic scattering and
breakup~\cite{GWHKG96,KRV99}, 2) $N$-$d$ radiative
capture~\cite{glocpt,Vea00}, 3) photodisintegration of $\tri$ and
$\het$~\cite{Wea00,EOL97}, and 4) electron-$\tri$ and
electron-$\het$ scattering~\cite{Vea00,gloelt} (see
Ref.~\cite{CS98} for a more complete list of references). The
calculations use a variety of approaches\cite{CS98}, such as the
Kohn variational principle, the Green Function Monte Carlo method,
or direct solution of the Faddeev equations, and have made it
possible to test our knowledge of the pairwise nucleon-nucleon (NN)
interaction and to study effects of possible three-nucleon (3N)
forces.

Calculations employing modern NN and 3N
interactions are capable of describing most of the experimental
results for the processes listed above.  However,
there is a well-known and large discrepancy
for the $N$-$d$ analyzing power at low energies.  What one
finds is that the predicted $A_y$ values are
sigificantly smaller in magnitude than the measurements
for both $p$-$d$ and $n$-$d$ elastic scattering.
Resolving this ``$A_y$ puzzle'' is a current and important
area of research~\cite{GWHKG96,KRTV96}.

In the present Letter, we extend some of the analyses described
above to the four-nucleon system.  We will present new accurate
computations for $p$-$\het$ elastic scattering at low energies.
We will also report new measurements of the analyzing power
for $p$-$\het$ scattering which have been obtained for the
purpose of testing whether the $A_y$ problem occurs in this
system as well.

Extending the accurate quantum calculations into the $4N$ system is
obviously of importance since it allows many new and stringent tests
of the nuclear interaction.
For example, accurate calculations of
the alpha particle binding energy $B_4$ have been
achieved recently~\cite{PPCPW97,Varga97,NKG00}, and it has been found
that the experimental value of the  binding energy
is reproduced with the same
NN and 3N potentials that fit the $\tri$ binding energy.
Since it appears that no
four-nucleon potential is necessary to explain the $\alpha $-particle
binding energy, one might
expect that NN + 3N interactions alone would also be sufficient to
describe the various four nucleon scattering processes.

The development of techniques for solving 4N problems is
also important for other reasons. Many reactions involving four
nucleons, like $d+d\ra
\heq+\gamma$ or $p+\het\ra \heq+e^+ +\nu_e$ (the $hep$ process), are of
extreme interest from the astrophysical point of view.  The
theoretical description of these processes constitutes a challenging
problem from the standpoint of nuclear few-body theory.  Its
difficulty can be appreciated by considering for example the $hep$
process.  In Ref.~\cite{Mea00} it was found that the capture
from the initial $P$-wave channels (``forbidden'' transitions) gives
about 40\% of the calculated $S$-factor, and this fraction depends
critically on the correct description of the dynamics of the
continuum and bound 4N states.

Moreover,  the study of 4N systems is important also for testing the
various many-body techniques developed for studying systems with
large ($\ge 4$) numbers of particles. These theories necessarily
include a number of approximations whose consequences can be
investigated by comparison with more sophisticated
calculations. The 4N system is the simplest system in which these checks
can be advantageously performed.

The new calculations reported in this Letter are based on
the Kohn variational principle
and make use of correlated hyperspherical harmonic (CHH) functions.
The same approach has already been applied successfully for a variety
of 3N processes~\cite{KRV99,KRV94}.
A previous application  for studying 4N
scattering at zero energy was already reported~\cite{KRV98}. In the
present paper, we have improved our calculation to determine P--
and D--wave phase shifts. The convergence of the P-wave shifts is rather
slow and has required a large technical effort to be achieved.
Results will be reported for calculations based on 1) the Argonne V18 
NN potential~\cite{AV18} (the AV18 model), and 2) the Argonne V18 NN
potential plus the Urbana IX 3N potential~\cite{PPCPW97} (the AV18UR
model). The present calculations represent the first attempt to study 
the effects of 3N forces on the $p$-$\het$ scattering
observables at energies greater than zero. 
In this paper we focus on low energies
where the convergence properties of our
theoretical approach are more satisfactory and where meaningful
comparisons with the experimental data can be performed.

The new measurements of the proton analyzing
power $A_y$ for $p$-$\het$ elastic scattering were obtained at
$E_{c.m.}=1.20$ and $1.69$ MeV.
The experiments were carried out at the University of
Wisconsin tandem accelerator laboratory.  Polarized protons from a
crossed-beam polarized ion source~\cite{PSV} were accelerated, momentum
analyzed by a $90{^\circ}$ bending magnet, and transported to a 1-m
scattering chamber.  The scattering chamber was filled with 43.4 Torr
of 99.95\% purity $^3$He gas, and was isolated from the beamline vacuum
by a $4.44 \times 10^{-5}$ cm thick Ni entrance foil located 1.27 cm from
the chamber center.

Elastically scattered protons were detected by three rectangular silicon
surface-barrier detectors, 60 to 100 $\mu$m thick, placed
symmetrically on each side of the scattering chamber.  A slit assembly
restricted the angular field of view to $\pm 0.34{^\circ}$.
The spectra were clean except for a small contaminant peak that was
well-separated from the peak of interest except at the most forward
angle.  At that angle, background subtraction was performed.

After passing through the scattering chamber, the beam entered a
polarimeter in which the beam polarization was determined using
$\vec p$-$\alpha$ elastic scattering~\cite{SCH}.  The polarimeter was
filled with one-half atmosphere natural He gas, and was separated
from the scattering chamber by a $2.54 \times 10^{-4}$ cm
thick Havar foil.  Because of the low beam energies, we could not
measure the beam polarization very accurately at the same
time as data were being taken.  However, previous experience indicates
that the beam polarization does not normally change significantly over
time.  Consequently, at least once a day we made a careful measurement
of the beam polarization by increasing the beam energy
to 4.0 MeV at the center of the polarimeter.  At
this energy, the polarimeter analyzing powers are known to 2\%.
Each such careful measurement of the beam polarization yielded a value
between 0.79 and 0.84 with typical statistical errors of~$\pm 0.01$.

The new measurements are shown in Fig.~\ref{fig:ay}.  The error bars include
statistical uncertainties and also at the extreme forward angle an
estimate of uncertainty in the background subtraction.  There is also
a scale factor uncertainty of 3\% due to beam polarization
uncertainties.

We turn now to the theoretical calculations.  Four-nucleon
scattering problems have been studied theoretically
for a  long time (see
Ref.~\cite{Tilley92} and references cited therein).
Recently, increases in computing power have opened the possibility
for accurate calculations of
the 4N observables using realistic models for NN
forces. These calculations have been performed mainly by means
of the Faddeev-Yakubovsky (FY) approach~\cite{CC98,Fonseca99}
and the Kohn variational principle~\cite{KRV98,Hofmann99}.
In this Letter, the wave functions of the scattering states
are expanded in terms of the CHH basis~\cite{KRV94} and the complex
form of the Kohn variational principle is applied~\cite{coko,K97}.

The wave function (WF) of a 4N state with total angular momentum $J$,
parity $\Pi$ and total isospin $T$, $T_z$ (in the present case
$T=T_z=1$) can be written as~\cite{KRV98}
\begin{equation}
  \Psi^{\qn}_{LS}=\Psi^\qn_C+\Phi^{\qn}_{LS}\ , \quad
  \Pi\equiv(-)^L\ .
   \label{eq:psitot}
\end{equation}
The first term $\Psi^\qn_C$ of Eq.~(\ref{eq:psitot}) must be
sufficiently flexible to guarantee a detailed  description in the
``internal region,'' where all the particles are close to each
other and the mutual interaction is large; $\Psi^\qn_C$ goes to
zero when the distance $r_i$ between the $\hel$ and the unbound proton
$i$ increases. This term in the WF is
expanded in terms of CHH basis functions, following the procedure
discussed in detail in Ref.~\cite{KRV95}.

The second term $\Phi^{\qn}_{LS}$
describes the asymptotic  configuration of the system, for large
$r_{i}$ values, where the nuclear $p$-$\hel$ interaction is
negligible.  Let us introduce the surface functions
\begin{equation}
   \Omega^{(\lambda)}_{LS\qn}= \sum_{i=1}^4
       \left\{ Y_L({\hat r}_i) \left[ \Psi^{\het}_{jk\ell}\;  \chi_i
       \right]_{S} \right\}_{JJ_z}
       {\cal R}^{(\lambda)}_L (r_i)
     \ ,\label{eq:surface1}
\end{equation}
where $\chi_i$ is  the spin function  of
the unbound nucleon $i$ and $\Psi^{\het}_{jk\ell}$ is
the $\hel$ bound state WF. This latter function is normalized
to unity and is  antisymmetric under the exchange of any pair of
particles $j$, $k$ and $\ell$. $\Psi^{\het}$ has been
determined as discussed in Ref.~\cite{KRV94} by using the  CHH
expansion for a three-body system.

The functions ${\cal R}^{(\lambda)}_L (r_i)$ of
Eq.~(\ref{eq:surface1}) are the ingoing ($\lambda\equiv -$) and
outgoing ($\lambda\equiv +$) radial solutions of the two-body
Schroedinger equation without nuclear interaction~\cite{K97}. The
asymptotic WF is then written as
\begin{equation}
   \Phi^{\qn}_{LS}= \Omega^{(-)}_{LS\qn} - \sum_{L'S'}
         {}^\qn{\cal S}^{SS'}_{LL'}\Omega^{(+)}_{L'S'\qn}\ ,
         \label{eq:asympt}
\end{equation}
where the S-matrix elements  ${}^\qn{\cal  S}^{SS'}_{LL'}$ give the
amplitude of the outgoing ($L'S'$) component relative to the
the ingoing ($LS$) wave.   The elastic S-matrix, whose
dimensionality is $1$ ($2$)  for the $J=0$ ($J>0$) states,
should be unitary since there are no open reaction channels
at the energies considered here.  It follows that
the eigenvalues of the S-matrix are written as $\exp(2{\rm i}
\delta^\qn_{LS})$, where
$\delta^\qn_{LS}$ is the eigenphase
shift of the ${}^{2S+1}L_{\qn}$ wave. These quantities
are calculated by means
of the complex form of the Kohn variational principle with a
procedure similar to that one used in the $N$-$d$
case~\cite{KRV94,K97}.

Note that with the present method the unitarity of the S-matrix
is assured only after the complete convergence of the CHH
expansion. Thus, for example, in the case of the $J=0$ waves, the
calculated S-matrix will be of the form ${}^\qnz{\cal
S}^{SS}_{LL}= \eta^{J\Pi}
\exp(2{\rm i} \delta^\qnz_{LS})$ where the  ``inelasticity'' parameter
$\eta^{J\Pi}$ may differ from 1 if the convergence is incomplete.
For $J>0$ states, the inelasticity parameter can be 
defined as $\eta^{J\Pi}=\sqrt{ {\rm Tr}[{\cal S}^\dag \cdot {\cal
S}]/2}$.

The  expansion of the internal part $\Psi^\qn_C$ is conveniently
studied by grouping the functions of the basis in ``channels''  (a
given channel contains CHH states with the same
angular-spin-isospin quantum numbers). The convergence of the $L=0$
waves ($J^\Pi=0^+$, $1^+$) at $E_{c.m.}=0$ was studied previously
in Ref.~\cite{KRV98}. At $E_{c.m.}=0$ and at the energies
considered here, a rather small number of channels is sufficient to
provide a good convergence. This is due mainly to the Pauli
principle which limits overlaps between the four nucleons. As a
consequence, the internal part is rather small and does not require
a large number of channels.

In contrast, for the $L=1$ waves ($J^\Pi=0^-$, $1^-$ and $2^-$)
the convergence rate is slow and many  channels have to
be included. In these cases, the interaction between the $p$ and
$\het$ clusters is very attractive (it has been speculated that
some 4N resonant states exist) and the construction of the
internal wave function is more delicate.

An example of convergence for the $0^-$ and $2^-$ states is
reported in Table~\ref{tab:con0m} (in the $J=2$ case the results
are relative to the $L=S=1$ wave). The calculation has been
performed using the AV18 potential at $E_{c.m.}=1.69$ MeV for a few
values of the number $N_c$ of channels included in the expansion of
$\Psi_C^J$. The values $\eta^{J\Pi}\approx 1$ for $N_c=0$ is
accidental, in fact, the value of $\eta^{J\Pi}$ increases after
including a few channels more. The convergence is reached only for
$N_c\gg 10$ when $|\eta^{J\Pi}-1|\approx 10^{-5}$.

At the energies considered here, the  scattering in the $L=2$ waves
($J^\Pi=1^+$, $2^+$ and $3^+$) is
very peripheral and the corresponding phase shifts are small. They can be
calculated with good precision by considering only the asymptotic part in
Eq.~(\ref{eq:surface1}).  The contribution of the waves with
$L>2$ is very tiny and has been disregarded.

The predicted analyzing powers are
compared with the measurements in Fig.~\ref{fig:ay}. 
The solid (dashed) curves
correspond to the AV18 (AV18UR) interaction model.
The main aspect to be seen in Fig.~\ref{fig:ay} is that the
calculations significantly underpredict the analyzing power by
approximately 30\% at $1.20$ MeV  and 40\% at $1.69 MeV$.
Similar results have previously been
seen and well documented for $N$-$d$ scattering.
We also see in Fig.~\ref{fig:ay}
that the 3N interaction has almost no effect on $A_y$.

In Table~\ref{tab:psa} the theoretical phase shift
parameters at $E_{c.m.}=1.20$  MeV are
compared  with those obtained in
the energy-dependent phase shift analysis (PSA) of
Ref.~\cite{AK93} which well reproduce the observables shown
in Fig.~\ref{fig:ay}. As can be seen, some phase shifts are
well reproduced by the theory. However, the ${}^3P_2$ and
${}^3P_1$ phase shifts are both sizeably underpredicted  by the theory.
For some partial waves the errors are too large to make a conclusive
statement about the quality of the calculation.

The $A_y$ observable is sensitive mainly to the $P$-wave phase shifts. At
$E_{c.m.}=1.20$ MeV, for example,
the $A_y$ maximum increases by approximately 42\% when the
${}^3P_2$ phase shift is changed by +10\%. On the other hand,
changing the ${}^3P_0$ (${}^3P_1$) phase by
+10\% decreases the $A_y$ maximum by 8 (6)\%. This observable
is in particular sensitive to the combination of phase shifts
$\Delta=0.5 [\delta({}^3P_1)+\delta({}^3P_2)]-\delta({}^3P_0)$. At
$E_{c.m.}=1.20$ MeV, the theoretical calculation  predicts
$\Delta=3.9^\circ $ ($3.8^\circ $)  with (without) the 3N force.
The corresponding experimental result (from the PSA)
is $\Delta=6.9^\circ \pm0.9 ^\circ$.

It is interesting to note that the discrepancy between the theoretical
and experimental $A_y$ is very much like that for the $N$-$d$
case. There the main problem is that the splitting between the
${}^4P_{1/2}$ phase and the average of the ${}^4P_{3/2}$ and ${}^4P_{5/2}$
phases is too small. For example, for $p$-$d$
scattering at $E_{c.m.}=2$ MeV, the calculations give
$ 0.5[\delta({}^4P_{3/2}) + \delta({}^4P_{5/2})] - \delta({}^4P_{1/2})
= 1.87^\circ $, whereas the phase shift fits to $A_y$ require
a  splitting of $2.61^\circ$~\cite{KRTV96}.
It is plausible to suspect that
the $A_y$ problems for $N$-$d$ and $p$-$\het$ both arise from the
same deficiency in the nuclear Hamiltonian.
Investigations in this direction are actively being pursued.

At the energies studied here old differential cross
section data exist~\cite{Fea54}. The theoretical
estimates are compared with these data in Fig.~\ref{fig:xs}.
Overall the agreement is good, especially in the minimum region, but at
large angles a
small discrepancy is seen. This discrepancy can once
again be traced to the underprediction of the calculated ${}^3P_2$ and
${}^3P_1$ phase shifts.
For this observable the inclusion of the 3N force produces small but
non-negligible effects. Finally, there are also some
measurements of the $\het$ analyzing
power in the energy range of interest~\cite{SSGK78}.
For this observable, the theoretical estimates are again below
the data, but unfortunately the experimental uncertainties are too
large to make a definitive statement.

In summary, we have performed
calculations of low-energy $p-\hel$ scattering observables,
employing a realistic interaction with
a 3N force, and including 
the effects of $P$ and $D$ waves.  A comparison with new
proton analyzing power measurements
demonstrates that the $A_y$ puzzle also exists
in the 4N system.

This work was supported in part by the National Science Foundation
under grant number PHY-9722554.

\begin{table}
\caption{Eigenphase shifts $\delta^J_{LS}$ (in degrees)  and inelasticity
parameters $\eta^{J\Pi}$ for the $p$-$\het$ scattering waves
${}^3P_0$ and ${}^3P_2$ at $E_{c.m.}=1.69$ MeV,  calculated with
the AV18 potential and the complex Kohn variational method. $N_c$
is the number of channels included in the CHH expansion of the wave
functions (the case $N_c=0$ corresponds to including in the WF only
the asymptotic part). }
\label{tab:con0m}
\begin{tabular}{cdd| cdd}
       \multicolumn{3}{c|}{${}^3P_0$ } &
       \multicolumn{3}{c}{${}^3P_2$ } \\
 $N_c$ & $\delta^0_{11}$ & $\eta^{0-}$  & $N_c$
       & $\delta^2_{11}$ & $\eta^{2-}$ \\
\hline
  0 &  3.9 & 1.0001 &0  & 8.4 & 1.0000\\
  2 &  4.2 & 1.0005 &2  & 8.9 & 1.0002\\
  4 &  4.8 & 1.0021 &5  & 10.0 & 1.0021\\
  6 &  6.3 & 1.0020 &10 & 12.4 & 1.0005\\
  8 &  6.5 & 1.0015 &15 & 12.9 & 1.0004\\
  9 &  6.6 & 1.0002 &25 & 13.3 & 1.0002\\
 18 &  6.9 & 1.0001 &35 & 13.5 & 1.0001\\
 45 &  7.0 & 1.0000 &100& 13.6 & 1.0000\\
\end{tabular}
\end{table}

\begin{table}
\caption{$S$- and $P$-wave phase shift and mixing angle parameters (in degrees)
at $E_{c.m.}=1.2$ MeV  calculated with the AV18 and the  AV18UR potentials.
The values obtained with the energy dependent PSA of
\protect\cite{AK93} are also shown.}
\label{tab:psa}
\begin{tabular}{cddd}
 wave  & AV18 & AV18UR & PSA \\
\hline
 ${}^1S_0$ & --33.3 & --31.3 & --27.4$\pm$3.5 \\
 ${}^3S_1$ & --28.8 & --27.1 & --26.5$\pm$0.6 \\
 ${}^3P_0$ & 4.1 & 3.2 & 2.6$\pm$0.6 \\
 ${}^3P_1$ & 8.1 & 7.4 & 10.1$\pm$0.5 \\
 ${}^3P_2$ & 7.7 & 6.9 & 8.9$\pm$0.5 \\
 ${}^1P_1$ & 6.5 & 5.5 & 4.2$\pm$1.5 \\
 $\epsilon(1^-)$ & --13.9 & --13.5 & --7.8$\pm$0.6 \\
\end{tabular}
\end{table}
\begin{figure}
\caption{Measurements of the proton analyzing power $A_y$
as a function of the scattering angle at $E_{c.m.}=1.20$ MeV (panel a)
and $1.69$ MeV (panel b). The  theoretical estimates obtained with the
AV18 (solid curves) and the AV18UR (dashed curves) interaction models
are also reported. }
\label{fig:ay}
\end{figure}

\begin{figure}
\caption{As in Fig.~\ref{fig:ay} but for the differential cross
section. The data
points are from Ref.\protect\cite{Fea54}. }
\label{fig:xs}
\end{figure}


\begin{references}

\bibitem{GWHKG96} W. Gl\" ockle  {\it et al.},
     Phys. Rep. {\bf 274}, 107 (1996).

\bibitem{KRV99}  A. Kievsky, S. Rosati and M. Viviani,
               Phys. Rev. Lett. {\bf 82}, 3759 (1999).

\bibitem{glocpt} J. Golak {\it et al.}, Phys. Rev C {\bf 62}, 054005 (2000).

\bibitem{Vea00} M. Viviani {\it et al.}, Phys. Rev {\bf C61}, 064003 (2000).

\bibitem{Wea00} E. Wulf {\it et al.}, Phys. Rev. {\bf C61}, 034003 (2000).

\bibitem{EOL97} V. D. Efros , W. Leidemann  and  G. Orlandini,
                Phys. Lett. {\bf B408}, 1 (1997).

\bibitem{gloelt}  S. Ishikawa {\it et al.}, Phys. Rev.
                  {\bf C57}, 39 (1998).

\bibitem{CS98} J. Carlson and R. Schiavilla, Rev. Mod. Phys. {\bf 70},
               743 (1998).

\bibitem{KRTV96} A. Kievsky, S. Rosati, W. Tornow and M. Viviani,
  Nucl. Phys. {\bf A607}, 402 (1996).

\bibitem{PPCPW97} B.S. Pudliner {\it et al.}, Phys. Rev. {\bf C56},
                  1720 (1997).

\bibitem{Varga97} K. Varga, Y. Ohbayashi and Y. Suzuki, Phys. Lett. {\bf
                  B396}, 1 (1997).

\bibitem{NKG00} A. Nogga, H. Kamada and  W. Gl\"ockle, preprint nucl--th/0004023.

\bibitem{Mea00} L. E. Marcucci {\it et al.}, Phys. Rev. Lett. {\bf
                84}, 5959 (2000).

\bibitem{KRV94} A. Kievsky, M. Viviani and S. Rosati, Nucl. Phys.
               {\bf A577}, 511 (1994).

\bibitem{AV18} R.B. Wiringa, V.G.J. Stoks, and R. Schiavilla,
               Phys. Rev. {\bf C51}, 38  (1995).

\bibitem{KRV98} M.\ Viviani, S.\ Rosati, and A.\ Kievsky,
                Phys.\ Rev.\ Lett.\ {\bf 81}, 1580 (1998).

%\bibitem{Giessen} H. Berg, W. Arnold, E. Huttel, H.H. Krause, J. Ulbricht,
%and G. Clausnitzer, Nucl. Phys. {\bf A334}, 21 (1980).

\bibitem{PSV} W. Haeberli {\it et al.}, Nucl. Instrum. Methods
{\bf 196}, 319 (1982).

\bibitem{SCH} P. Schwandt, T. Clegg, and W. Haeberli, Nucl. Phys.
{\bf A163}, 432 (1971).

\bibitem{Tilley92} D. R. Tilley, H.R. Weller and G.M. Hale,
                   Nucl. Phys. {\bf A541}, 1 (1992).

\bibitem{CC98} F. Ciesielski and J. Carbonell, Phys. Rev. {\bf C58}, 58
              (1998).

\bibitem{Fonseca99} A. C. Fonseca, Phys. Rev. Lett. {\bf 83},
                   4021 (1999).

\bibitem{Hofmann99} H. M. Hofmann and G. M. Hale, Nucl. Phys. {\bf
                    A613}, 69 (1997).

\bibitem{coko} B.I. Schneider and T.N. Rescigno, Phys. Rev. {\bf A37},
                 3749 (1988).

\bibitem{K97} A. Kievsky, Nuclear Physics {\bf A624}, 125 (1997).

\bibitem{KRV95} M. Viviani, A. Kievsky and S. Rosati,  Few--Body Systems
               {\bf 18}, 25  (1995).

%\bibitem{KRV4N} M. Viviani, A. Kievsky and S. Rosati,
%               in preparation

\bibitem{AK93} M.T. Alley and L. D. Knutson, Phys. Rev. {\bf C48},
                  1901 (1993).

\bibitem{Fea54} K. F. Famularo {\it et al.}, Phys. Rev. {\bf 93}, 928
                (1954).

\bibitem{SSGK78} G. Szaloky {\it et al.}, Nucl. Phys. {\bf A303}, 51
(1978).

\end{references}
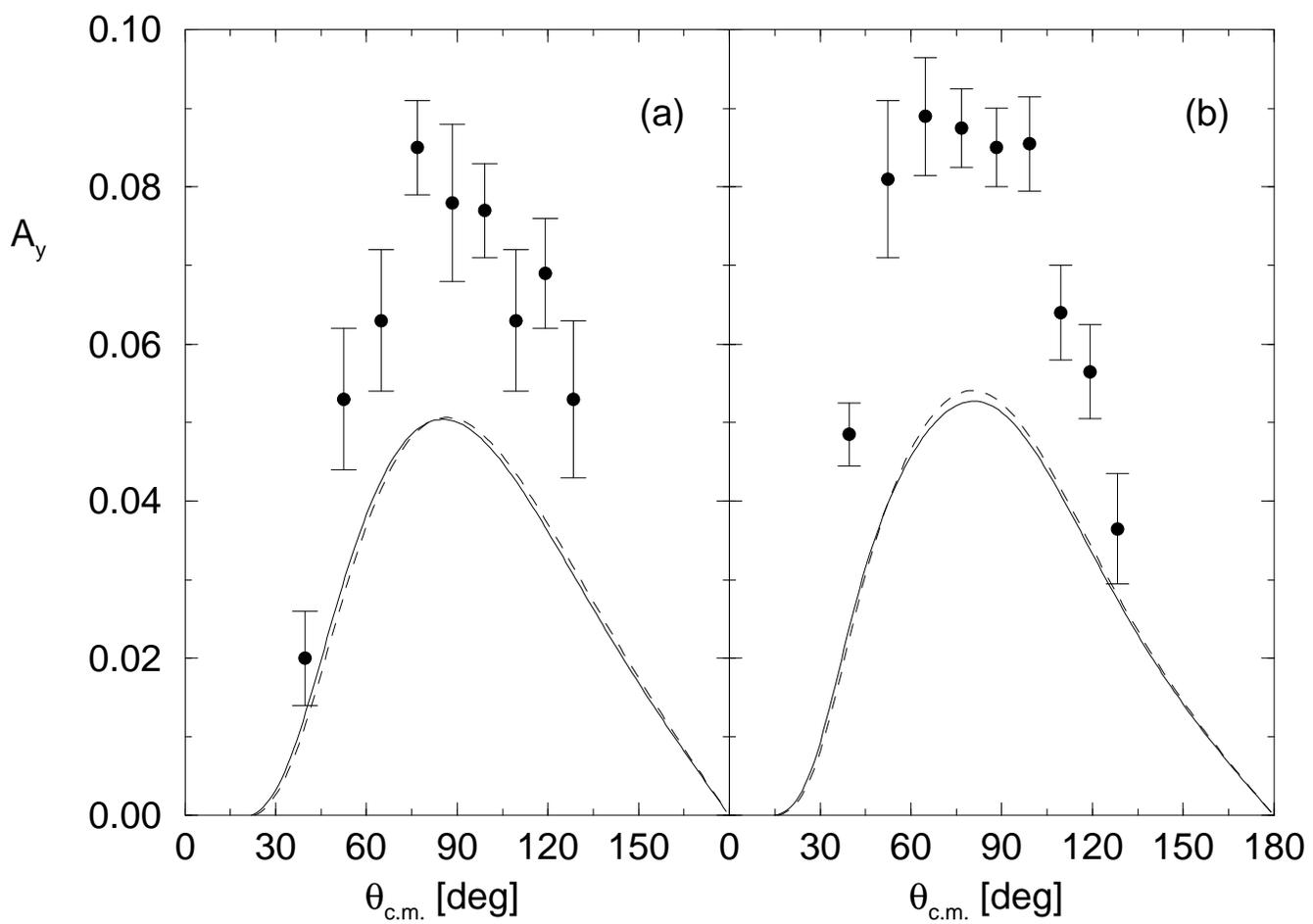
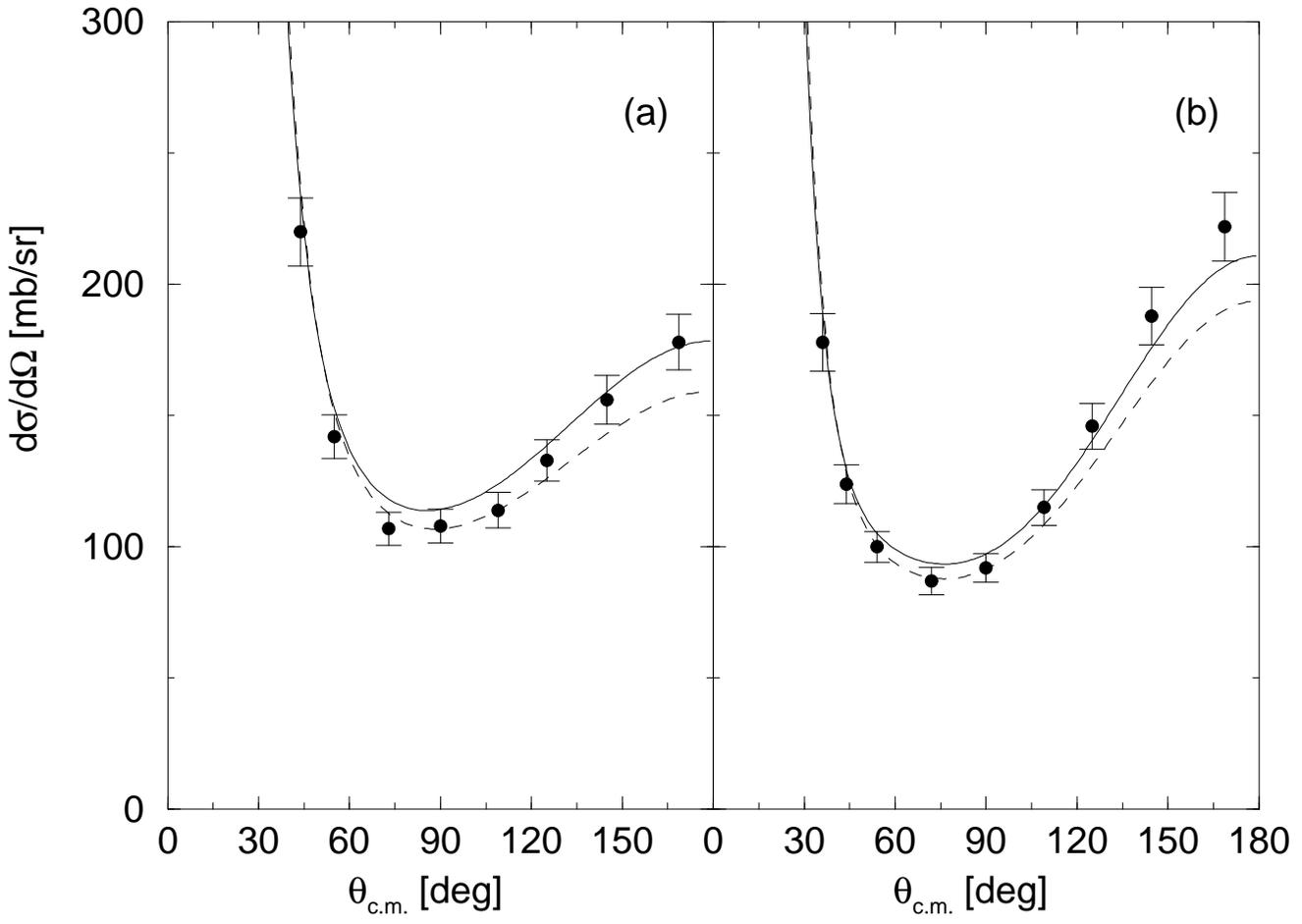
\end{document}